 \documentclass[12pt,preprint]{aastex}

\shorttitle{SOLAR MODELS} \shortauthors{Yang,W. M. \& Bi, S. L.}

\begin{document}


\title{SOLAR MODELS WITH REVISED ABUNDANCES AND OPACITIES}

\author{W. M. Yang\altaffilmark{1}}
\affil{National Astronomical Observatories/Yunnan Observatory,
Chinese Academy of Sciences, Kunming 650011, China.}
\email{yangwuming@ynao.ac.cn}

\and
\author{S. L. Bi\altaffilmark{2}}
\affil{Department of Astronomy, Beijing Normal University, Beijing
100875, China.}
\email {bisl@bnu.edu.cn}

\altaffiltext{1}{Graduate School of the Chinese Academy of
Sciences.} \altaffiltext{2}{National Astronomical
Observatories/Yunnan Observatory, Chinese Academy of Sciences.}

\begin{abstract}
Using reconstructed opacities, we construct solar models with low
heavy-element abundance. Rotational mixing and enhanced diffusion of
helium and heavy elements are used to reconcile the recently
observed abundances with helioseismology. The sound speed and
density of models where the relative and absolute diffusion
coefficients for helium and heavy elements have been increased agree
with seismically inferred values at better than the 0.005 and 0.02
fractional level respectively. However, the surface helium abundance
of the enhanced diffusion model is too low. The low helium problem
in the enhanced diffusion model can be solved to a great extent by
rotational mixing. The surface helium and the convection zone depth
of rotating model M04R3, which has a surface Z of 0.0154, agree with
the seismic results at the levels of 1 $\sigma$ and 3 $\sigma$
respectively. M04R3 is almost as good as the standard model M98.
Some discrepancies between the models constructed in accord with the
new element abundances and seismic constraints can be solved
individually, but it seems difficult to resolve them as a whole
scenario.
\end{abstract}


\section{Introduction}
Recent analyses of solar photospheric abundance have shown a
significant reduction of the abundances of C, N, O, and other heavy
elements \citep{lod03, asp04, all05, asp05, sco06}. As a result, the
ratio of the heavy-element abundance to hydrogen abundance, Z/X, is
reduced from 0.023 \citep[hereafter GS98]{gre98} to 0.0171
\citep[hereafter AGS]{asp04} (or 0.0177 \citet{lod03}); and the
surface Z of the Sun decreases from 0.017 to 0.0126 (or 0.0133 ).
Many solar models constructed in accord with this low Z disagree
with the seismically inferred sound speed and density profiles,
convection zone (CZ) depth, and helium abundance of the CZ
\citep{bas04a, bas04b, mon04, turck04, bah05, guz05}.

\citet[hereafter BA04]{bas04a} and \citet{bah05} found that the 15 -
20 per cent increase of OPAL opacities at the base of the CZ can
resolve the discrepancies between the solar models with low Z and
helioseismology. However, \citet{sea04} and \citet{bad05} showed
that the increase in the opacities is no more than 2.5 per cent near
the base of the CZ. \citet{ant05} and \citet{bah05} found that
increasing neon abundance can solve the discrepancies. However,
\citet{sch05} and \citet{you05} showed that the ratio of Ne/O is
consistent with the value given by AGS. Another attempt to resolve
the problem focuses on increasing the rates of diffusion of helium
and heavy elements. Models with enhanced diffusion rates are in
better agreement with helioseismology than the model without the
enhanced diffusion. However, the sound speed and the density are
still far from the seismic results, or the surface Z/X value is
still too high and the helium abundance of the CZ is too low
\citep{bas04a, mon04, guz05}.

Using helioseismic data, \citet{ant06} determined that the solar
metal abundance is 0.0172 $\pm$ 0.002. This is consistent with that
of GS98. \citet{del06} found that both Fe/H and O/H are more
consistent with the  values of GS98 than the  values of AGS.
\citet{ayr06} also found that the oxygen abundance is close to the
value of GS98. If these results are further confirmed, the problems
induced by low Z will disappear. However, \citet{asp06} argued that
the AGS results are trustworthy. Recently, \citet{sco06} also found
that the low carbon abundance is in good agreement with the findings
based on entirely different indicators of AGS and with the values
determined by \citet{ire06} from lunar grains irradiated by solar
wind. \citet{sco06} claimed that their results are more reliable
than those of \citet{ayr06}. The answer to the problems is still an
open question.

Our aim is to construct a solar model, using recently determined
abundances, which can agree with seismic constraints. We apply
straight multipliers to the diffusion velocity to enhance the rates
of element diffusion. Although the theoretical error of
gravitational settling rate is of the order of about 15 per cent
\citep{tho94}, there are some significant uncertanties in the
treatment of the element diffusion \citep{guz05}. Our multipliers of
the diffusion coefficients are very high, despite the fact that
there is no obvious physical justification for such high
multipliers, as has been pointed out by BA04 and \citet{guz05}.
However, the multipliers are required in our models for diffusion to
reduce the heavy-element abundance of the CZ from the GS98 value to
near the AGS value. In order to get the same helium in a rotating
model as in a non-rotating model, a multiplier of the element
diffusion is required because rotational mixing reduces the degree
of gravitational settling. Other enhanced diffusion-rate models have
been discussed by BA04, \citet{mon04}, and \citet{guz05}. The main
difference between our models and those of others is that we include
rotational effects in order to resolve the low helium problem of the
CZ. We compare our results with those of BA04 and \citet{guz05} in
Table \ref{tab1}.

\section{Solar models }
\subsection{Properties of our solar models}
We use the Yale Rotation Evolution Code (YREC7) to construct our
solar models. However, we modify the code to include two tables of
OPAL EOS for the Z diffusion; and we correct an error in rotating
model in the calculation of the chemical compositions. We use
reconstructed OPAL opacities \citep{igl96} and new low-temperature
opacities \citep{fer05}, both with the GS98 or AGS mixtures. In
principle, the EOS table should also be reconstructed using AGS
mixtures, but it has been found that the effects of the change of
heavy elements on the EOS can be ignored \citep{bas04a, bah04,
bah05, guz05}. We therefore use the OPAL EOS \citep{rog96} in all
our models. Element diffusion is included for helium and metals
\citep{tho94}. Energy transfer by convection is treated according to
the standard mixing-length theory, and the boundaries of the
convection zones are determined by the Schwarzschild criterion. We
take the solar age to be 4.57 Gyr. Luminosity $L_{\odot}$ =
3.8418$\times10^{33}$ erg/s, and radius $R_{\odot}$ = 6.9598$\times
10^{10}$ cm.

We construct the following six models: 1) M98, a standard model with
GS98 mixture opacities; 2) M04, a model with AGS mixture opacities;
3) M04D, same as M04 but enhancing the element diffusion
\citep{tho94}; 4) M04R1, same as M04D but with rotation
\citep{pin89} and diffusion coefficient \citep{zahn93} added for
shear instability; 5) M04R2 and 6) M04R3, both the same as M04R1 but
with different diffusion multipliers, as shown in Table \ref{tab1}.

Some of the parameters of the models are summarized in Table
\ref{tab1}. The mixing-length parameter $\alpha$, $Z_{init}$ and
$Y_{init}$ are free parameters adjusted to obtain the observed solar
radius, luminosity and surface element abundances. Finally,
$R_{cz}/R_{\odot}$, $(Z/X)_{s}$, $Y_{s}$, and $Z_{s}$ are the
results of calculations at the age of 4.57 Gyr.

\subsection{Results}
\begin{figure}
\includegraphics[angle=-90,scale=0.50]{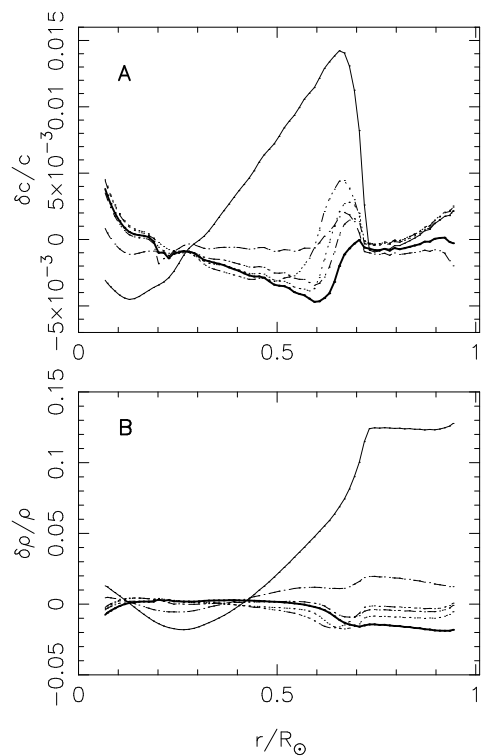}
\caption{A) Sound speed difference between the Sun and the model. B)
Density difference between the Sun and the model. The solar values
were derived from MDI data \citep{bas00}. The long-dashed line
refers to M98. The solid line indicates M04. The dash-dot-dot-dot
line shows the results of M04D. The dotted line is given by M04R1.
The dash-dot line and the bold solid line are corresponding to M04R2
and M04R3 respectively. \label{fig1}}
\end{figure}

Some of the calculation results are showed in Table \ref{tab1}. The
base of the CZ is at 0.7335 $R_{\odot}$ for M04 and is 20 $\sigma$
different from the seismically inferred 0.713 $\pm$ 0.001
$R_{\odot}$ \citep{bas97}. The surface helium abundance of 0.2294 is
6 $\sigma$ away from the seismically inferred value 0.2485 $\pm$
0.0034 (BA04). The sound speed and density differences between M04
and the Sun are shown in Figure \ref{fig1}, where those of the Sun
were given by \citet{bas00}. M04 is in strong disagreement with the
seismically inferred sound speed and density profiles, the depth of
the CZ, and envelope helium abundance.

\citet{asp04} suggested that increased diffusion might be able to
resolve these disagreements. Thus, we construct the model M04D by
multiplying the diffusion coefficients for helium and heavy elements
by factors of 2.4 and 3.8 respectively. However, we have no physical
justification for these multipliers. This method was firstly
proposed by \citet{guz05}. The base of the CZ of M04D is at 0.7168
$R_{\odot}$, which agrees with the seismically inferred value
\citep{bas97} at about the 3 $\sigma$ level. The sound speed and
density of M04D are better consistent with the seismic data than
that of M04. The difference of the sound speed and density between
M04D and the Sun, $\delta c/c$ and $\delta\rho/\rho$, is less than
0.005 and 0.02 respectively. These values are close to those of the
standard model M98. In fact, the density profile of M04D is even
slightly better than that of M98. However, the surface helium
abundance of 0.2225 is too low, which disagrees with the seismically
inferred value 0.2485 (BA04) at the level of 8 $\sigma$.

The rotational mixing can reduce the degree of gravitational
settling \citep{cha95, yan06}. We thus construct a rotating model,
M04R1, to study the low helium problem of M04D. The rotational
mixing in all our models is treated as a diffusion process
\citep{pin89}; and the rotational mixing processes include Eddington
circulation, Goldreich-Schubert-Fricke instability, and the secular
shear instability \citep{zahn93}. In all our rotating models, the
surface rotation rate is about 2.86$\times 10^{-6}$ rad/s at the age
of 4.57 Gyr. The sound speed and density differences  between M04R1
and the Sun are shown in Figure \ref{fig1}. They are almost the same
as those in M04D. The surface helium abundance of M04R1 is 0.2368,
which is 3 $\sigma$ away from 0.2485 (BA04) but agrees with that of
\citet{kos97}. However, the base of the CZ is at 0.7206 $R_{\odot}$
and disagrees with the seismically inferred value at about the 7
$\sigma$ level.

In order to get a model which is more consistent with seismic
constraints than M04R1, we relax the constraint of the heavy
elements determined by AGS and construct models M04R2 with Z/X =
0.01976 and M04R3 with Z/X = 0.0208. The sound speed and density
differences between M04R2 and the Sun are less than 0.004 and 0.01
respectively; and the surface helium abundance of M04R2 is 0.2454
within the constraint of observation (BA04). However, the depth of
the CZ in M04R2 disagrees with the seismically inferred value at
about the 7 $\sigma$ level. The $\delta c/c$ and $\delta\rho/\rho$
of M04R3 are less than 0.005 and 0.02 respectively. The surface
helium abundance and the base position of the CZ in M04R3 agree with
the seismically inferred values at the levels of 1 $\sigma$ and 3
$\sigma$ respectively; but the envelope Z of 0.0154 is higher than
the value of AGS. This indicates that a model with high heavy
elements is more consistent with seismic constraints than the model
with AGS heavy elements.

In the FULL1M of BA04, the differences $\delta c/c$ and $\delta
\rho/\rho$ are larger than 0.01 and 0.05 respectively. In the
\citet{guz05} models, the $\delta c/c$ is larger than 0.005. We
compare our results with these models in Table \ref{tab1}.

\section{Discussion and conclusions}
In this study, we assume that the diffusion coefficients for helium
and heavy elements can be enhanced respectively. Thus the
discrepancies between the models with the low Z and
helioseismological theory can be reduced, but the helium abundance
of the CZ maintains too low in such models. In enhanced diffusion
models, there is a gradient of element abundances caused by element
settling at the base of the CZ. The rotational secular shear is
highly sensitive to the gradients of mean molecular weight
\citep{pin89}, and the rotational mixing can smooth out the gradient
of the element abundances. The settling of the helium abundance can
thus be partly counteracted by the rotational mixing, and the
surface helium abundance of the rotating model can thus be enhanced.

After relaxing the constraint of the Z/X in model M04R3, the
sound-speed profile, density profile, depth of the CZ, and surface
helium abundance are almost the same as those in the standard model
M98; but the surface Z of 0.0154 is higher than the value of AGS.

The difference in the sound speed and density in our models can be
improved by enhancing the diffusion coefficients for helium and
heavy elements. Thus, the maximum difference in sound speed
decreases from 0.014 in M04 to 0.0045 in M04D and M04R1. The density
profiles of the enhanced diffusion models are even slightly better
than that of M98. The surface helium abundance of 0.2225 in model
M04D is too low and is 8 $\sigma$ different from that of BA04. The
surface helium abundance of the rotating model M04R1 is 0.2368,
which agrees with that of \citet{lod03} and \citet{kos97}; but the
position of the base of the CZ in this model is 7 $\sigma$ away from
the seismically inferred position \citep{bas97}. The surface helium
abundance and the base position of the CZ in M04R3 are in agreement
with seismic results at the levels of 1 $\sigma$ and 3 $\sigma$
respectively; but the surface heavy-element abundance of 0.0154 is
higher than the value of AGS. Some problems between the new element
abundances and helioseismological theory can be solved individually,
but it seems difficult to resolve them as a whole scenario.

\acknowledgments We are grateful for help of Daniel Kister, useful
remarks from the anonymous referee, and the support by the NSFC
through projects 10473021 and 10433030.

\clearpage

\clearpage

\begin{table}
\begin{center}
\caption{Model parameters. \label{tab1}}
\begin{tabular}{ccccccccc}

\tableline\tableline

Model & $Y_{init}$ & $Z_{init}$ &  $\alpha$&  Multiplier
& $R_{cz}$/$R_{\odot}$ & $(Z/X)_{s}$ & $Y_{s}$ & $Z_{s}$ \\
\tableline M98  & 0.2794   & 0.0201  & 2.166  & 1.0\tablenotemark{a}
(1.0)\tablenotemark{b}& 0.7151\tablenotemark{c}  & 0.0247 & 0.2487 & 0.01809 \\
M04  & 0.26066  & 0.0148  & 1.608  & 1.0 (1.0) & 0.7335  & 0.0174 & 0.2294 &0.01322  \\
M04D & 0.28618 & 0.019619 & 1.7827  & 2.4 (3.8) & 0.7168  & 0.0176 & 0.2225  &0.01347 \\
M04R1 & 0.285803 & 0.019497 & 1.71799& 2.4 (3.8) & 0.7206  & 0.0177 & 0.2368 &0.01330  \\
M04R2 & 0.2836 & 0.018892 & 1.67555& 2.0 (2.5) & 0.7206  & 0.01976 & 0.2454 &0.01462  \\
M04R3 & 0.28358 & 0.01886 & 1.6889& 2.0 (2.0) & 0.7167  & 0.0208 & 0.2450 &0.015396  \\
\tableline \tableline
Basu FULL1M &   &   &  & 1.65& 0.7233  & 0.0171 & 0.2244 &0.0130 \\
Basu FULL2M &   &   &  & 1.65& 0.7138  & 0.0218 & 0.2317 &0.01639 \\
Guzik model3 &  &   &  &3    & 0.7022  & 0.0196 &0.1926 &0.01552  \\
Guzik model5 &  &   & & 1.5; 4& 0.7175  & 0.0206 & 0.2269 &0.01561
\\\tableline
\end{tabular}
\tablenotetext{a}{ The multiplier for the diffusion coefficient of
the helium;} \tablenotetext{b}{ The multiplier for the diffusion
coefficient of the heavy elements;} \tablenotetext{c}{ Using OPAL
EOS96, \citet{bah04} obtained $R_{cz}$ = 0.7155 $R_{\odot}$.}

\end{center}
\end{table}

\end{document}